\documentclass[aps,prb,twocolumn]{revtex4}

\usepackage{amsmath,amssymb,mathrsfs}
\usepackage{amsbsy,amsthm,amsfonts}
\usepackage{color}
\usepackage[colorlinks,citecolor=blue]{hyperref}
\usepackage{bm}
\usepackage{graphics,graphicx}

\usepackage{color}

\newcommand\figref[1]{Fig.~\ref{#1}}

\newcommand\sectref[1]{Sec.~\ref{#1}}
\newcommand\eqnref[1]{(\ref{#1})}

\newcommand{\bfa}   {\mathbf{a}}

\newcommand{\bfb}   {\mathbf{b}}
\newcommand{\bfB}   {\mathbf{B}}

\newcommand{\bfd}   {\mathbf{d}}
\newcommand{\bfD}   {\mathbf{D}}

\newcommand{\bfe}   {\mathbf{e}}
\newcommand{\bfE}   {\mathbf{E}}

\newcommand{\bfh}   {\mathbf{h}}
\newcommand{\bfH}   {\mathbf{H}}
\newcommand{\bfJ}   {\mathbf{J}}
\newcommand{\bfM}   {\mathbf{M}}
\newcommand{\bfP}   {\mathbf{P}}
\newcommand{\bfj}   {\mathbf{j}}
\newcommand{\bfk}   {\mathbf{k}}
\newcommand{\bfm}   {\mathbf{m}}

\newcommand{\calM}  {\mathcal{M}}
\newcommand{\calS}  {\mathcal{S}}

\newcommand{\bfq}   {\mathbf{q}}

\newcommand{\bfr}   {\mathbf{r}}

\newcommand{\iu}  {i}  
\newcommand{\K}  {k}  
\newcommand{\Omegarm}   {\mathrm{\Omega}} 
\newcommand{\thinc}   {\theta_\mathrm{inc}}

\newcommand{\eZM}   {e_{\mathrm{ZM}}}

\begin{document}

\title[Nonasymptotic Homogenization: Uncertainty Principles]
{Nonasymptotic Homogenization of Periodic Electromagnetic Structures:\\
Uncertainty Principles}

\author{Igor Tsukerman}
\email{igor@uakron.edu}
\affiliation{Department of Electrical and Computer Engineering, The
  University of Akron, OH 44325-3904, USA}

\author{Vadim A. Markel}
\thanks{On leave from the Department of Radiology, University of Pennsylvania, 
Philadelphia, Pennsylvania 19104, USA}
\email{vmarkel@mail.med.upenn.edu; vmarkel@fresnel.fr}
\affiliation{Aix-Marseille Universit\'{e}, CNRS, Centrale Marseille, 
Institut Fresnel UMR 7249, 13013 Marseille, France}

\date{\today}
\begin{abstract}
We show that artificial magnetism of periodic dielectric or metal/dielectric structures
has limitations and is subject to at least two ``uncertainty principles''. 
First, the stronger the magnetic response (the deviation of the effective permeability tensor from identity), 
the less accurate (``certain'') the predictions of any homogeneous model. 
Second, if the magnetic response is strong, then homogenization cannot accurately reproduce the
transmission and reflection parameters and, \textit{simultaneously},
power dissipation in the material. 
These principles are general and not confined to any particular method of homogenization.
Our theoretical analysis is supplemented with a numerical example: a hexahedral lattice of cylindrical air holes
in a dielectric host. Even though this case is highly isotropic, which might
be thought as conducive to homogenization, the uncertainty principles remain valid.
\end{abstract}


\maketitle

\section{Introduction}
\label{sec:Intro}
\subsection{Overview}

Over the last 15--20 years, artificial magnetism of periodic dielectric or metal/dielectric structures 
(``metamaterials'' and photonic crystals) has attracted much attention and is often tacitly assumed 
to have no principal limitations, especially in the ideal case of negligible losses.
In this paper, however, we argue that such limitations do exist. Namely, the stronger the magnetic response 
(as measured by the deviation of the optimal effective permeability
tensor from identity), the less accurate (``certain'') predictions of any homogeneous model
of the material are. We call this an \textit{uncertainty  principle} (UP)
for the effective parameters of metamaterials. It should be emphasized
that this principle constitutes a general limitation and is not confined 
to any particular method of homogenization.

We also introduce another uncertainty principle: if the magnetic response 
of a periodic structure is strong,
then homogenization cannot accurately reproduce the
TR (transmission and reflection) parameters and, \textit{simultaneously},
power dissipation (the heating rate) in the material. 
This ultimately follows from the fact that the TR coefficients
are governed by the \textit{boundary values} of Bloch waves in the material,
while power is related to the \textit{volume average} of a quadratic function
of that wave.  

The premise of our analysis is that the objective of homogenization is to predict,
as accurately as possible, transmission and reflection of waves
by a periodic electromagnetic structure -- for simplicity,
a slab (this eliminates complications due to edges and corners).
We also consider a \textit{homogeneous} slab of the same thickness and
with a material tensor $\calM$ such that the TR
coefficients for the original and homogeneous slabs agree to a given level
of tolerance over a sufficiently broad range of illumination conditions.

Our analysis is general and does not depend on a particular homogenization theory
and on the way the effective tensor $\calM$ is determined.
We consider periodic electromagnetic structures
	in the framework of classical electrodynamics.
	This does include plasmonic metamaterials with feature sizes above $\sim$10--20~nm,
	when classical (frequency-dependent) permittivity $\epsilon(\omega)$  is still applicable.
	However, the field of metamaterials is currently so broad\cite{Sihvola07}
	that our theory may not be directly applicable to some types (superconducting materials, magnonic materials~\cite{Mikhaylovskiy10,Zivieri12}, etc.).
	

We argue that for TR to be rendered accurately, not only the dispersion
relation in the bulk, but also the boundary conditions on the surface of
the slab must be approximated well. 
%
It then follows that the amplitudes of the Bloch waves within the material
are dictated by an accurate boundary match with the incident and reflected waves.
Loosely speaking, this boundary match fixes the wave impedance, 
while the Bloch number fixes the dispersion relation. Both pieces of information 
are necessary to define the effective material tensor unambiguously.

A critical question then is whether the resulting tensor is (or could be)
independent of the angle of incidence.\footnote{Evanescent waves can be included if ``angle''
is understood as an imaginary quantity.} Clearly, angular-dependent material parameters
do not have their traditional meaning, and their practical utility is limited.

An illumination-independent tensor certainly exists in the classical
homogenization limit, when the ratio of the lattice cell size $a$
to the vacuum wavelength $\lambda$ approaches zero 
\cite{Bakhvalov-Panasenko89,Bensoussan78,Milton02}.
However, magnetic effects vanish in that limit \cite{Bossavit05,Tsukerman08},
and therefore this case is not of primary interest to us here. In
the remainder, we shall assume a \textit{non}-asymptotic regime, where 
%
%
$a$ and $\lambda$ are of
the same order of magnitude ($0.1 \lesssim a / \lambda \lesssim 0.5$).
Then, in general, there is an appreciable surface wave whose behavior is quite involved. 
In his previous work \cite{Markel-Schotland12}, one of the co-authors showed
that surface waves have zero averages of the tangential field components
on the interface boundary. Therefore they do not affect coarse-scale
boundary conditions and homogenization, although they of course do contribute
to the near-field behavior.

\textit{Remark 1}. If lattice periods along the interface
are greater than the vacuum wavelength, then surface waves are propagating 
rather than evanescent. We exclude this case from consideration.

Surface waves propagating along interfaces can carry energy and, 
for a sample finite in all directions, can be reflected off its edges
and scattered off its corners. This cannot be addressed in the simplified setup
adopted in this paper: a slab with a finite thickness but infinite in the
remaining two directions. However, since surface waves do not generally exist
in homogeneous media (expect for special circumstances such as total internal
reflection or surface plasmons at interfaces where Re~$\epsilon$ changes its sign),
it is clear that the presence of such waves in periodic structures and reflection
of these waves from the edges can only be detrimental to homogenization and cannot weaken the
uncertainty principles presented here.
Thus, to fix key ideas, we disregard surface waves. The interested reader may find
further information in Refs.~\onlinecite{Markel-Schotland12,Xiong13,Xiong15},
although research on this subject is still far from being complete.

%
There is ample evidence in the existing literature that effective parameters 
of metamaterials may have limited accuracy and validity.
As an instructive example\cite{Simovski-review09},
the celebrated negative-index metamaterial due to Smith \textit{et al}.
\cite{Smith00} cannot be homogenized for a range of wavelengths in the vicinity of the second
$\Gamma$-point, even though these wavelengths are relatively long
($a / \lambda \sim 0.1$).
Another notable example is the work of the Jena and Lyngby groups \cite{Menzel10},
who show that high symmetry of a metamaterial cell does not imply 
optical isotropy, especially in frequency ranges where the effective
index is negative.

In our previous publications \cite{Markel-Tsukerman-PRB2013,Tsukerman-Markel14}, 
we brought to the fore an interplay between magnetic response,
the accuracy of homogenization and the range of illumination conditions.
Here we extend this line of reasoning and show that
not only negative index, but a strong magnetic response must
unfortunately be accompanied by lower accuracy of effective parameters,
unless illumination conditions are restricted to a narrow range.

\subsection{Local vs. Nonlocal Parameters}

This paper deals exclusively with \textit{local} effective material parameters.
Because of extensive discussions of nonlocality (or ``spatial dispersion'')
in the literature on metamaterials, it might be tempting
to draw a connection between the uncertainty principles of this paper and
nonlocality; hence brief comments on the latter are called for.

In classical electromagnetism, a local linear material relationship
(say, between the $\bfD$ and $\bfE$ fields and with magnetoelectric coupling
ignored for the sake of brevity)
has the form $\bfD(\bfr) = \epsilon(\bfr) \bfE(\bfr)$ -- that is,
one field at any given point is related to another field at that same point.
In contrast, a classical nonlocal relationship is usually written as
\begin{equation}
\label{eqn:D-eq-Eps-star-E}
   \bfD(\bfr) \,=\, \int_{\Omegarm} \mathcal{E}(\bfr, \bfr') \, \bfE(\bfr') \, d \bfr'
\end{equation}
where $\mathcal{E}$ is a convolution kernel and $\Omegarm$ is the region
occupied by the material in question.

However, even if the nonlocal relation \eqnref{eqn:D-eq-Eps-star-E} could be rigorously
established in the bulk, it would require special treatment at interfaces due to the lack of
translational invariance. We are not aware of any theory that would rigorously
define the kernel $\mathcal{E}(\bfr, \bfr')$ as a function of two position vectors
near the metamaterial/air interface. Moreover,
it is not clear how such a kernel could be put to practical use, as
all metamaterial devices proposed so far depend critically on a local description 
of the effective medium. 

If, for the argument's sake, one were to accept the view that \textit{weak} spatial dispersion
is equivalent to \textit{local} parameters $(\epsilon, \mu)$ 
(even though we have argued against this view\cite{Markel-Tsukerman-PRB2013}),
then our requirement that parameters be local would still be justified.

In the remainder of the paper, we lay out theoretical arguments supporting 
the two uncertainty principles summarized above and present an instructive example:
a triangular lattice of cylindrical air holes
in a dielectric host, as investigated previously by Pei \& Huang \cite{Pei12}.
This example is interesting because, despite a high level of isotropy
around the $\Gamma$-point in the second photonic band,
which may be thought as conducive to homogenization,
the uncertainty principles remain valid.

\section{Formulation of the problem}
\label{sec:Formulation}
%
The formulation of the homogenization problem was given in
Ref.~\onlinecite{Tsukerman-Markel14}. For completeness,
we include it here in a shortened form, omitting some technical details
not critical for the analysis in this paper.
	
We consider homogenization of periodic composites characterized by the
intrinsic permittivity and permeability $\tilde{\epsilon}(\bfr)$ and
$\tilde{\mu}({\bf r}) = 1$. The effective parameters will be denoted
with $\epsilon$ and $\mu$ (without the tilde).  The composite constituents
are assumed to be linear, local and intrinsically nonmagnetic so that
$\tilde{\mu}({\bf r}) = 1$ everywhere in space. Also, we assume that
$\tilde{\epsilon}(\bfr)$ is a scalar (a multiple of the identity
tensor). In contrast, the effective parameters $\epsilon$ and $\mu$
can be different from unity and are, generally, second-rank tensors.

The tilde sign is used for all lattice-periodic quantities. For
example, Bloch-periodic functions (Bloch waves) are written in the
form
\begin{equation*}
f(\bfr) = \tilde{f}(\bfr) \exp(i \bfq \cdot \bfr) \ ,
\end{equation*}
\noindent
where $\bfq$ is the Bloch wave vector. Here symbol $\bfq$ is used
to distinguish the Bloch wave vector of a given medium from a generic
wave vector $\bfk$.
In the case of orthorhombic lattices, periodicity is expressed as
\begin{equation}
\label{periodicity}
\tilde{f}(x + n_x a_x, y + n_ya_y, z + n_za_z) = \tilde{f}(x,y,z) \ ,
\end{equation}
\noindent
where $a_{x,y,z}$ are the lattice periods and $n_{x,y,z}$ are arbitrary integers. 
Of course, \eqref{periodicity} is assumed to hold only if both points $\bfr=(x,y,z)$ and
$\bfr^\prime = (x + n_x a_x, y + n_ya_y, z + n_za_z)$ are simultaneously located
either inside the composite or in a vacuum.

Fine-level fields -- that is, the exact solutions to the macroscopic
Maxwell's equations -- are denoted with small letters $\bfe$, $\bfd$,
$\bfh$ and $\bfb$. Capital letters $\bfE$, $\bfD$, $\bfH$, $\bfB$ are
used for coarse-level fields that would exist in an equivalent
effective medium, still to be defined. 
The constitutive relations for the fine-level fields are
\begin{equation*}
   \bfd(\bfr) = \tilde{\epsilon} (\bfr)  \bfe(\bfr) \ , \ \
   \bfb(\bfr) =                          \bfh(\bfr) \ .
\end{equation*}
\noindent
Note that $\bfh(\bfr) = \bfb(\bfr)$ because the medium is assumed to
be intrinsically nonmagnetic. 

Our analysis is in the frequency domain with the $\exp(-i \omega
t)$ phasor convention. At a working frequency $\omega$, the
free-space wave number $\K$ and wavelength $\lambda$ are
\begin{equation*}
   \K = \frac{\omega}{c} = \frac{2\pi}{\lambda}
\end{equation*}
%

We compare transmission and reflection of electromagnetic waves through/from 
(separately) two slabs of a thickness $d$ each, for simplicity infinite in the longitudinal direction
(half-space  can be viewed as a valid particular case $d \rightarrow \infty$). 
One of these slabs is composed of a given metamaterial (i.e. has a periodic structure),
while the other one contains a homogeneous medium with a yet unknown
material tensor $\calM$. To any (monochromatic) plane wave incident on the surface
$z = 0$ of either slab at an angle $\thinc$, there correspond transmission (T) and
reflection (R) coefficients $R_{\mathrm{MM}}$, $R_{\mathrm{hmg}}$; $T_{\mathrm{MM}}$, $T_{\mathrm{hmg}}$,
where subscripts `MM' and `hmg' indicate the metamaterial and homogeneous cases,
respectively. Under the condition of Remark 1, reflection and transmission
coefficients for the metamaterial slab are well defined.

The difference between the reflection and transmission 
coefficients produced by the two slabs  
will be referred to as the \textit{TR-discrepancy} $\delta_{TR}$:
$$
   \delta_{TR} \,\equiv\,  \| R_{\mathrm{hmg}}(\thinc) - R_{\mathrm{MM}}(\thinc) \| 
$$
\begin{equation}\label{eqn:TR-discrepancy}
   +\,
   \| T_{\mathrm{hmg}}(\thinc) - T_{\mathrm{MM}}(\thinc) \|
\end{equation}
where $\| \cdot \|$ is a desired norm -- say, the $L_2$-norm
over a given range of illumination conditions, e.g. $\thinc \in [-\pi/2, \pi/2]$
if all propagating waves but no evanescent ones are considered.
Our analysis below applies to any homogenization theory that produces a tensor $\calM$ 
approximately minimizing the TR-discrepancy $\delta_{TR}$.

%
\section{First Uncertainty Principle: Magnetic Response vs Accuracy of Homogenization}
\label{sec:Magnetic-response-vs-accuracy}
%
As noted in the Introduction, magnetic characteristics of metamaterials become trivial
in the zero-cell-size limit \cite{Sjoberg05-149,Bossavit05,Tsukerman08} (assuming that the intrinsic
material parameters remain bounded). Thus a strong magnetic response can only be achieved
if the cell size forms an appreciable fraction of the vacuum wavelength.
It is the objective of this section to show that stronger
effective magnetic properties are unavoidably accompanied by 
lower approximation accuracy of the metamaterial by a homogeneous medium 
with local parameters. We call this an
``uncertainty principle'' (UP) of local homogenization.

Although parts of our analysis are similar to those of Ref.~\onlinecite{Tsukerman-Markel14},
we do \textit{not} assume that the $\calM$ tensor has been determined using necessarily
the procedure of Ref.~\onlinecite{Tsukerman-Markel14}. Rather, let $\calM$ be found using
\textit{any} method (say, parameter retrieval as the most common example).

To avoid unnecessary mathematical complications and to keep our focus
on the physical essence of the problem, we present our analysis of the UP for the $s$-mode
(the $\bfE$ field in the $z$ direction, the $\bfH$ field in the $xy$-plane),
with a plane wave impinging in the $xy$-plane on a half-space filled with a metamaterial, 
i.e. a dielectric structure characterized by a permittivity $\epsilon(\bfr)$ periodic in 
the $x$ and $y$ directions with the same (for simplicity) lattice constant $a$. 

We introduce 
normal $n$ and tangential $\tau$ coordinates relative to the material/air interface $S$;
$n$ points from the air (on the side of the incident wave) into the metamaterial.

%

A Bloch wave with a wavevector $\bfq$ is
\begin{equation}\label{eqn:e-Bloch-wave}
  e_B(\bfr, \bfq) ~=~ \tilde{e}_B(\bfr) \exp(\iu \bfq \cdot \bfr) 
\end{equation}
where subscript `B' indicates a Bloch wave-related
quantity.
The tangential component of the respective $\bfh$-field is
\begin{equation}\label{eqn:h-Bloch-wave}
  h(\bfr) \,=\, \frac{1}{\iu \K} \frac{\partial e}{\partial n}
  \,=\, 
  \tilde{h}_{B}(\bfr) \exp(\iu \bfq \cdot \bfr),
\end{equation}
(only the tangential component is used in the analysis, and therefore
subscript `$\tau$' is dropped for brevity of notation). The periodic factor 
for the magnetic field is
\begin{equation}\label{eqn:tilde-h-via-e}
  \tilde{h}_{B}(\bfr) \,=\, \frac{q_n}{\K} \, \, \tilde{e}_B(\bfr) + 
  \frac{1}{\iu \K} \frac{\partial \tilde{e}_B(\bfr)}{\partial n}
\end{equation}
We now compare wave propagation from the air into a half-space
filled with 

\noindent
(i) a metamaterial, and 

\noindent
(ii) a homogeneous medium, with its material
tensor yet to be determined to minimize the TR-discrepancy.
\footnote{Transmission is easy to define only for a finite-thickness slab;
for a half-space, the focus is on reflection as a function of the angle of incidence.}

In both cases (i) and (ii), the field in the air is given by
\begin{equation}\label{eqn:e-air}
  e_{\mathrm{air}}(\bfr) ~=~ E_{\mathrm{inc}} \left[ \exp(\iu \bfk_{\mathrm{inc}} \cdot \bfr) 
  \,+\, R \exp(\iu \bfk_{\mathrm{r}} \cdot \bfr) \right]
\end{equation}
\begin{equation}\label{eqn:htau-air}
  h_{\mathrm{air}}(\bfr) \,=\,  E_{\mathrm{inc}} \cos \thinc 
  \left( \exp(\iu \bfk_{\mathrm{inc}} \cdot \bfr) 
    \,-\, R \exp(\iu \bfk_{\mathrm{r}} \cdot \bfr) \right)
\end{equation}
where the tangential component is again implied for $\bfh$.

It will be convenient to assume that the reflection coefficient $R$
is exactly the same in cases (i) and (ii). 
Strictly speaking, there can be
(and in practice will be) some approximation tolerance; however,
introducing this tolerance explicitly would obscure the analysis while adding little
to its physical substance.

As already stated, in the case of a metamaterial we ignore the surface wave. 
Then the field in the metamaterial is
just the Bloch wave \eqnref{eqn:e-Bloch-wave}, \eqnref{eqn:h-Bloch-wave}.
This should be compared with the transmitted wave in the homogeneous half-space:
\begin{equation}\label{eqn:eT-homog}
  E_{T}(\bfr) ~=~ E_{T0} \exp(\iu \bfk_{T} \cdot \bfr) 
\end{equation}
\begin{equation}\label{eqn:hT-homog}
  H_{T}(\bfr) \,=\,  H_{T0} \exp(\iu \bfk_T \cdot \bfr) 
\end{equation}
Phase matching between \eqnref{eqn:eT-homog} 
and \eqnref{eqn:e-Bloch-wave} implies that for best approximation one must have
\begin{equation}\label{eqn:k-eq-q}
   \bfk_{T} \,=\, \bfq
\end{equation}
Further, due to the boundary conditions at the material/air interface,
the amplitudes of the transmitted wave in the equivalent homogenized medium
must be
\begin{equation}\label{eqn:ET0-eq-avrg-EB}
   E_{T0} \,=\, \langle \tilde{e}_B \rangle_S ~~~~~~(=\, (1+R) E_{\mathrm{inc}})
\end{equation}
\begin{equation}\label{eqn:HT0-eq-avrg-EB}
   H_{T 0} \,=\, \langle \tilde{h}_B \rangle_S 
\end{equation}
where $\langle, \cdot, \rangle_S$ denotes the air/cell boundary average.
Indeed, if, say, condition \eqnref{eqn:ET0-eq-avrg-EB} were to be violated,
then in the homogenized case there would be a spurious jump  
of the $E$-field across the air/material interface, with the \textit{nonzero mean}
$$
   E_{T0} - \langle \tilde{e}_B \rangle_S = E_{T0} - (1+R) E_{\mathrm{inc}}
$$
(assuming that the interface boundary is at $n = 0$). This jump will result in
a commensurate far-field error.

\textit{Remark}. We require that boundary conditions hold in the sense of averages
\eqnref{eqn:ET0-eq-avrg-EB}, \eqnref{eqn:HT0-eq-avrg-EB} rather than point-wise
because zero-mean discrepancies between a Bloch wave and a plane wave at
the boundary are unavoidable. Indeed, the Bloch wave in an inhomogeneous medium
has higher-order spatial harmonics that cannot be matched by a plane wave.
Conditions \eqnref{eqn:ET0-eq-avrg-EB}, \eqnref{eqn:HT0-eq-avrg-EB} ensure
that the discrepancy between the Bloch field on the material side and
plane waves on the air side affect only the near field, as long as $a < \lambda$.

Now that the field amplitudes in the homogenized material have been determined,
we can find the material tensor for which the dispersion relation (in essence, Maxwell's equations)
will be satisfied. We'll be primarily interested in the case of four-fold ($C_4$ group)
symmetry, which is particularly instructive. (In a more general situation, the effective
tensor needs to be defined via ensemble averages, as was done in our previous paper
\cite{Tsukerman-Markel14}.) For $C_4$ cells, the material tensor is diagonal
(in particular, there is no magnetoelectric coupling) and, moreover,
$\mu_{\tau \tau} = \mu_{nn}$. In the remainder, we shall focus on the $\mu_{\tau \tau}$
entry of the tensor.

Maxwell's $\nabla \times \bfE$-equation for the generalized plane wave 
\eqnref{eqn:eT-homog}, \eqnref{eqn:hT-homog} gives the amplitude of the 
tangential component of the $\bfB$-field in this wave:
\begin{equation}\label{eqn:BT0-eq-kE}
   B_{T0} \,=\, \frac{1}{\iu \K} \, k_{Tn} \,  E_{T0} 
\end{equation}
or, substituting $\bfk_{T} = \bfq$ from \eqnref{eqn:k-eq-q} and
$E_{T0}$ from \eqnref{eqn:ET0-eq-avrg-EB},
\begin{equation}\label{eqn:BT0-eq-q-eB-mean}
   B_{T 0} \,=\, \frac{q_{n}}{\K} \,  \langle \tilde{e}_B \rangle_S 
\end{equation}
This, along with Eq. \eqnref{eqn:HT0-eq-avrg-EB} for the amplitude of $H_{T}$,
leads to the following expression for the effective magnetic permeability:
\begin{equation}\label{eqn:mu-tau-eq-Btau-over-Htau}
   \mu_{\tau \tau} \,=\, \frac{B_{T 0}}{H_{T 0}}
   \,=\, \frac{q_{n} \langle \tilde{e}_B \rangle_S}{\K \langle \tilde{h}_{B} \rangle_S}
   \,=\, \frac{q_{n} \langle \tilde{e}_B \rangle_S}
   {q_{n} \langle \tilde{e}_B \rangle_S - \iu \langle \partial_n \tilde{e}_B \rangle_S}
\end{equation}
where we inserted expression \eqnref{eqn:tilde-h-via-e} for $\tilde{h}_{B}$.
Switching for algebraic convenience from permeability to reluctivity, we arrive at the following
surprisingly simple expression:
\begin{equation}\label{eqn:zeta-tau-via-etilde-dn}
   \zeta_{\tau \tau} \,\equiv\, 1 - \mu_{\tau \tau}^{-1} \,=\, 
   \frac{\iu \, \langle \partial_n \tilde{e}_B \rangle_S} 
   {q_{n} \langle \tilde{e}_B \rangle_S} \,=\, 
   \frac{\iu \, \langle \partial_n \tilde{e}_B \rangle_S} 
   {q \cos \theta_B \langle \tilde{e}_B \rangle_S}
\end{equation}
where $\theta_B$ is the propagation angle for the Bloch wave.
It is instructive to split $\tilde{e}_B$ in \eqnref{eqn:zeta-tau-via-etilde-dn}
into its mean value $e_0$ and zero-mean $\eZM$,
$$
    \tilde{e}_B \, \equiv \, e_0 \,+\, \eZM,
    ~~~ e_0 = \mathrm{const}, ~~ \int_C \eZM \, dC = 0
$$
Then \eqnref{eqn:zeta-tau-via-etilde-dn} becomes
\begin{equation}\label{eqn:zeta-tau-via-e-zeromean}
   \zeta_{\tau \tau} \,=\, 
   \frac{\iu \, \langle \partial_n \eZM \rangle_S} 
   {q \cos \theta_B (e_0 \,+\,\langle  \eZM \rangle_S)}
\end{equation}
It becomes immediately clear that magnetic effects in metamaterials
are due entirely to higher-order spatial harmonics of the Bloch wave,
manifested in $\eZM$. (If  $\eZM = 0$, the Bloch wave is just a plane wave,
and $\zeta_{\tau \tau} = 0$.) To avoid any misunderstanding,
note that $\eZM$ by definition has a zero average in the \textit{volume} of the cell
but in general not on its surface, which makes all the difference 
in \eqref{eqn:zeta-tau-via-e-zeromean}.

The behavior of Bloch waves in inhomogeneous lattice cells is complicated,
and there are no simple closed-form expressions for these waves. (Approximations 
are well known but exist only as formal solutions of large
linear systems developed in a finite basis, e.g. in a plane-wave basis.)
From the qualitative physical perspective, however, one may conclude that,
due to the complex dependence of $\eZM$ on $\theta_B$ (i.e. on the illumination conditions),
$\zeta_{\tau \tau}$ is in general angle-dependent. Moreover, this angular dependence 
will tend to be stronger when the magnetic effects (nonzero $\zeta_{\tau \tau}$) 
are themselves stronger, as both are controlled by $\eZM$.
This conclusion can also be supported quantitatively (see Appendix A)
but does not have the status of a mathematical theorem; the door is therefore
still open for engineering design and optimization, with a compromise between
the strength of magnetic response and homogenization accuracy.

\section{Second Uncertainty Principle: TR-discrepancy vs. Power Discrepancy}
\label{sec:TR-discrepancy-vs-power-discrepancy}
\begin{figure}
\centering
\includegraphics[width=0.85\linewidth]{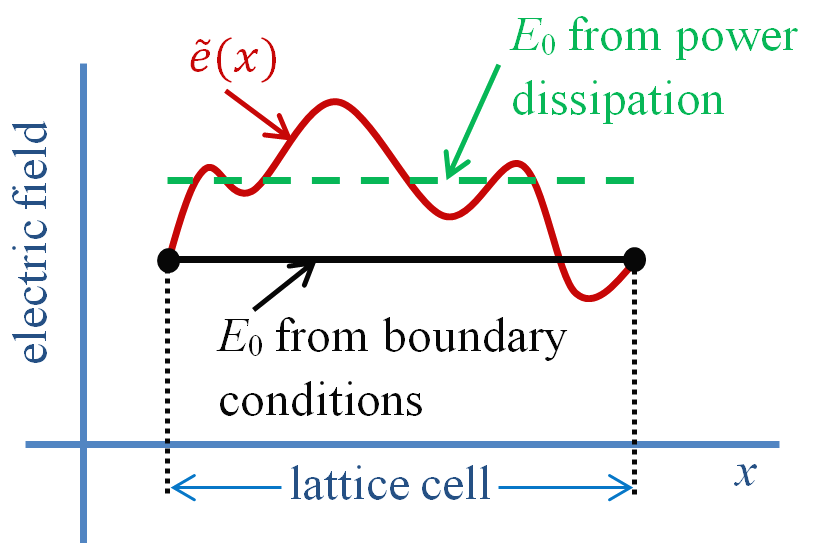}
\caption{Coarse-level amplitudes derived from power dissipation and from
boundary conditions are in general different.}
\label{fig:E0-from-bc-and-power}
\end{figure}

%
In this section, we put forward a second uncertainty principle:
if magnetic response is appreciable, homogenization cannot accurately reproduce
both TR \textit{and} power dissipation (the heating rate).
The root cause of that can be easily grasped from the simplified 1D sketch in
\figref{fig:E0-from-bc-and-power}. Let the periodic factor $\tilde{e}(x)$
of a Bloch wave in a given lattice cell be approximated, in a homogenized medium,
by a plane wave of amplitude $E_0$. If it is power dissipation in the homogenized medium 
that is matched to the actual power,
then $E_0$ should be at the root-mean-square (rms) level indicated by the dashed line.\footnote{
The term `rms' here should be taken literally only in special cases
such as 1D or $s$-modes; otherwise power averaging is more complicated.}
On the other hand, if it is the boundary conditions that are matched (which is necessary
for rendering the TR correctly), then $E_0$ must have a different
value, indicated by the solid line in the figure.

Let us proceed to a more formal analysis.
In the homogenized case, the general expression for the total current density
within the medium is, in the frequency domain,
\begin{equation}\label{eqn:J-eq-dtP-curl-M} 
	\bfJ \,=\, -\iu \omega \bfP \,+\, c \nabla \times \bfM
\end{equation}
where
$$
	\bfP = \frac{1}{4 \pi} (\bfD - \bfE) = \frac{1}{4 \pi} (-\frac{1}{\iu \K} \nabla \times \bfH - \bfE)
$$
\begin{equation}\label{eqn:P-eq-D-minus-E} 
  = -\frac{1}{4 \pi \K} \left(\bfq \times \bfH + \K \bfE \right)
\end{equation}
$$
	\bfM = \frac{1}{4 \pi} (\bfB - \bfH) = \frac{1}{4 \pi} \left(\frac{1}{\iu \K} \nabla \times \bfE - \bfH \right)
$$
\begin{equation}\label{eqn:M-eq-B-minus-H} 
  = \frac{1}{4 \pi \K} \left(\bfq \times \bfE - \K \bfH \right)
\end{equation}
%
Substituting expressions for $\bfP$ and $\bfM$ \eqnref{eqn:P-eq-D-minus-E} and \eqnref{eqn:M-eq-B-minus-H}
into the expression for $\bfJ$ \eqnref{eqn:J-eq-dtP-curl-M}, one obtains
$$
  \bfJ \,=\,  \frac{\iu c}{4 \pi} \left(\bfq \times \bfH + \K \bfE \right)
   \,+\, \frac{c}{4 \pi}  \nabla \times \left(\frac{1}{\K} \bfq \times \bfE - \bfH \right)
$$
$$
  =\,  \frac{\iu \omega}{4 \pi} \, \bfE
   \,+\, \frac{\iu c}{4 \pi \K} \, \bfq \times ( \bfq \times \bfE)
$$
$$
  =\,  \frac{\iu \omega}{4 \pi} \, \bfE
   \,+\, \frac{\iu c}{4 \pi \K} \, \left( \bfq (\bfq \cdot \bfE) - (\bfq \cdot \bfq) \bfE \right)
$$
Expressing the heating rate as\footnote{Although different
and not equivalent expressions for the heating rate exist \cite{Markel08},
our final conclusion here does not depend on which of these expressions
is used in the analysis.} $\frac12 \mathrm{Re} (\bfJ \cdot \bfE^*)$,
one obtains per-cell power dissipation in the homogenized medium\cite{Markel08}
\begin{equation}\label{eqn:power-homogenized} 
  W_C \,=\, -\frac{c}{8 \pi \K} \, \mathrm{Im}
  \int_C
  \left\{ (\bfq \cdot \bfE) (\bfq \cdot \bfE^*) - (\bfq \cdot \bfq) (\bfE \cdot \bfE^*) \right\} dC
\end{equation}
%

The power calculation for the actual fine-scale fields is similar
but simpler, since for intrinsically nonmagnetic components magnetization
$\bfm$ is by definition zero. Hence we have
$$
  \bfj \,=\,  \frac{\iu \omega}{4 \pi} \left(\frac{1}{\iu \K} \nabla \times \bfh + \bfe \right)
  =\,  \frac{\iu \omega}{4 \pi} \left(\frac{1}{\K^2} \, \bfq \times (\bfq \times \bfe) + \bfe \right)
$$
$$
  \,=\, \frac{\iu \omega}{4 \pi \K^2} \left( \bfq (\bfq \cdot \bfe) 
  + (\K^2 - (\bfq \cdot \bfq)) \, \bfe \right)
$$
Thus the actual power dissipation per lattice cell is
\begin{equation}\label{eqn:power-actual} 
  w_C \,=\, - \frac{c}{8 \pi \K}
  \mathrm{Im} \left\{
  \int_C
  \left[ (\bfq \cdot \bfe) (\bfq \cdot \bfe^*) 
    - (\bfq \cdot \bfq) \, (\bfe \cdot \bfe^*) \right] \right\}  dC
\end{equation}
Let us consider an $s$-wave 
as a simple but representative model; the conclusion generalizes
to arbitrary waves in 2D or 3D periodic structures. For an $s$-wave,
the ``longitudinal'' $\bfq \cdot$ terms vanish, as the electric field is, by definition,
orthogonal to $\bfq$. Then, for power dissipation $W_C$ on the coarse level
to be equal to the actual power $w_C$, the amplitude $E_0$ on the coarse level
has to satisfy
\begin{equation}\label{eqn:E0-sq-eq-int-e2} 
  |E_0|^2 V_C \,=\,  \int_C | \tilde{e} |^2  dC
\end{equation}
This follows from the direct comparison of expressions \eqnref{eqn:power-homogenized},
\eqnref{eqn:power-actual} and the fact that $\bfE$ and $\bfe$ contain the same
Bloch exponential: $E(\bfr) = E_0 \exp (\iu \bfq \cdot \bfr)$, 
$e(\bfr) = \tilde{e}(\bfr) \exp (\iu \bfq \cdot \bfr)$.

On the other hand, to represent TR accurately, one needs to honor the boundary conditions (see \sectref{sec:Magnetic-response-vs-accuracy}). Thus, according to \eqnref{eqn:ET0-eq-avrg-EB},
\begin{equation}\label{eqn:E0-eq-avrg-EB}
   E_0 \,=\, \langle \tilde{e}_B \rangle_S
   \,=\, e_0 + \langle \eZM \rangle_S
\end{equation}
If \eqnref{eqn:E0-sq-eq-int-e2} were to hold with $E_0$ satisfying \eqnref{eqn:E0-eq-avrg-EB},
one would have
$$
  |e_0 + \langle \eZM \rangle_S|^2 \, V_C \,=\,  \int_C | e_0 + \eZM |^2  dC
$$
and the right hand side simplifies because the zero-mean function $\eZM$
is orthogonal to the constant $e_0$:
\begin{equation}\label{eqn:Esim-sq-eq-int-esim2} 
  |\langle \eZM \rangle_S|^2 + 2 \mathrm{Re} 
  \{ \langle \eZM \rangle_S \, e_0^* \} \,=\, 
  \langle |\eZM |^2 \rangle_C
\end{equation}
Since the volume distribution of a Bloch mode and the respective value of $e_0$ 
are only loosely related to its boundary values, the above condition 
is quite restrictive and cannot be expected to
hold for any given wave, let alone for \textit{all} Bloch waves,
traveling in different directions or evanescent.
A trivial exception is $\eZM \equiv 0$, in which
case the Bloch wave turns into a plane wave and there are no magnetic effects.
The stronger these effects, the more strongly \eqnref{eqn:Esim-sq-eq-int-esim2}
will in general be violated.


\section{Numerical Example: the Uncertainty Principle for a Problem with High Isotropy}
\label{sec:Results-Pei-huang-triang-lattce}
%
\subsection{The Setup}\label{sec:Triangular-lattice-setup}
%
As an instructive example, we consider the hexahedral lattice of cylindrical air holes
in a dielectric host investigated previously by Pei \& Huang \cite{Pei12}. 
%
The radius of each air hole is $r_\mathrm{cyl} = 0.42a$, the dielectric permittivity 
of the host is $\epsilon_{\mathrm{host}} = 12.25$;  $s$-polarization (TM-mode, one-component
$E$ field perpendicular to the plane of the figure).
This example is interesting because in the second photonic band
it exhibits a high level of isotropy around the $\Gamma$-point
and a negative effective index.

The elementary cell of this lattice can also be viewed as a rhombus,
with the corresponding real-space lattice vectors 
$$
   \bfa_1 = a \hat{x};  ~~~
   \bfa_2 = \frac{a}{2} \, \left( \hat{x} + \sqrt{3} \, \hat{y} \right)
$$
and reciprocal vectors
$$
    \bfb_1 = \kappa  \left(1, \, -\frac{1}{\sqrt{3}} \right); ~~~
    \bfb_2 = \kappa  \left(0, \, \frac{2}{\sqrt{3}} \right),   ~~~ \kappa \equiv \frac{2 \pi}{a}
$$
The real and reciprocal vectors satisfy the standard Kronecker-delta relation
\begin{equation}
  \bfa_{\alpha} \cdot \bfb_{\beta} \,=\, 2\pi \delta_{\alpha \beta},
  ~~~ \alpha , \beta = 1,2
\end{equation}

\begin{figure}
\centering
\includegraphics[width=0.99\linewidth]{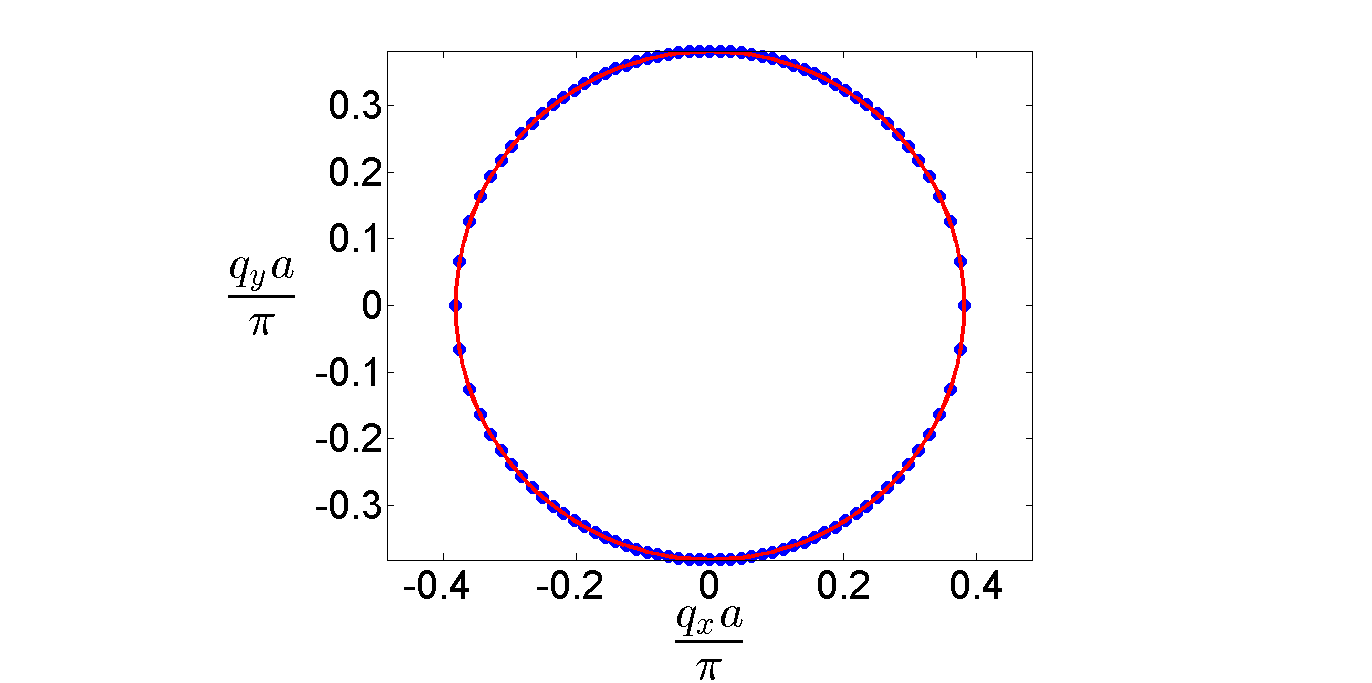}
\caption{
An almost circular first-Brillouin-zone isofrequency contour 
for the Pei-Huang\cite{Pei12} triangular lattice of air holes
$r_\mathrm{cyl} = 0.42a$, $\epsilon_{\mathrm{host}} = 12.25$, 
with $a = 0.365 \lambda$ near the $\Gamma$-point $a / \lambda \approx 0.368$.
Markers: numerical data points; solid line: circular fit.
The isotropy of the dispersion relation is evident and has been noted by Pei \& Huang.}
\label{fig:circular-isofreq-triang-lattice-a0365}
\end{figure}

\begin{figure}
\centering
\includegraphics[width=0.99\linewidth]{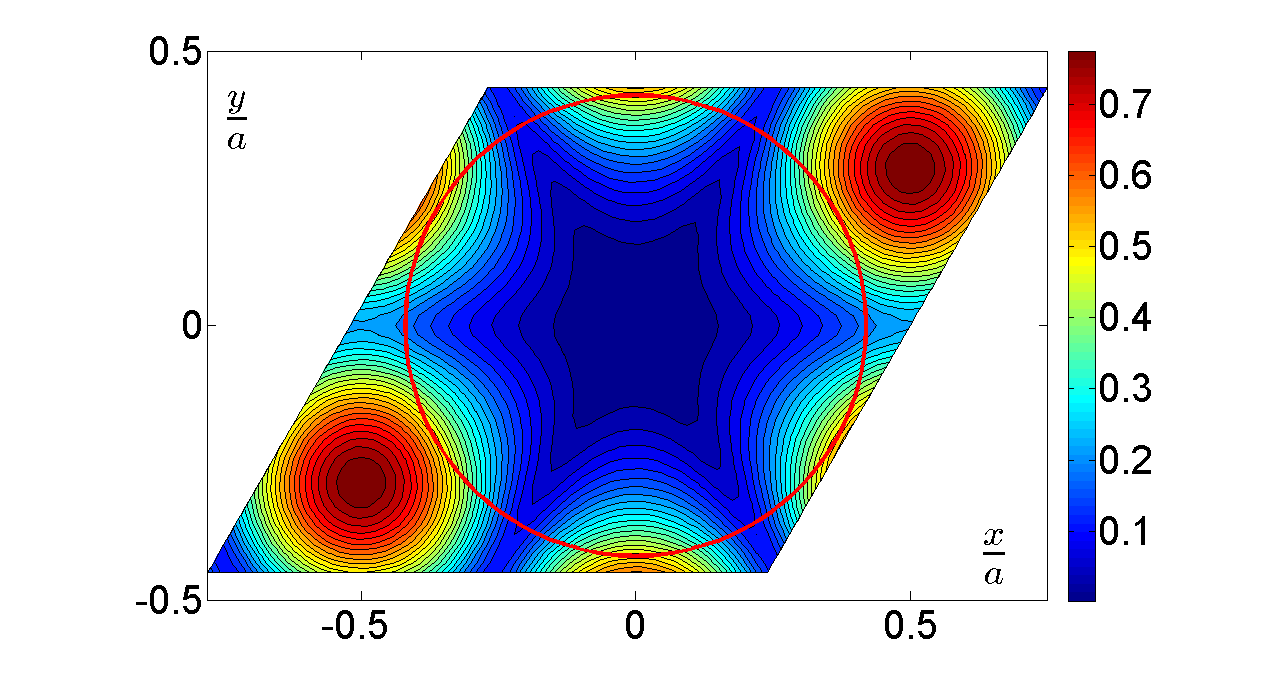}
\includegraphics[width=0.99\linewidth]{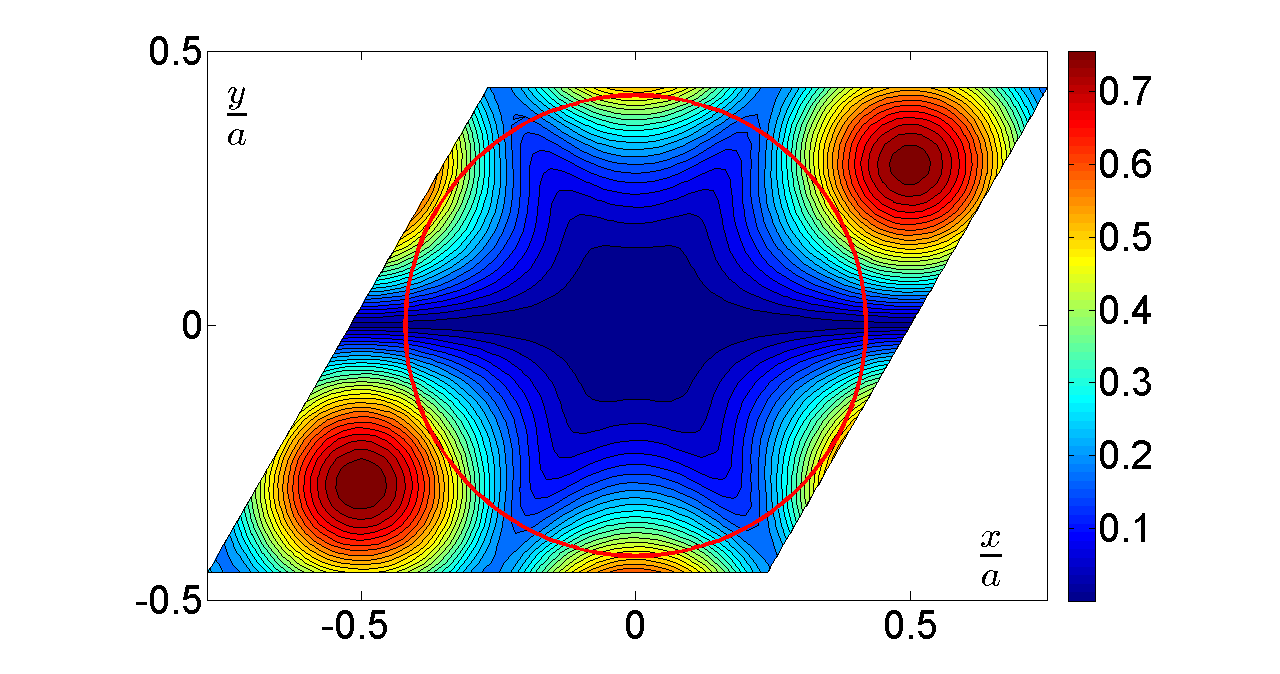}
\caption{The absolute value of the electric field for the $s$-mode
with $q_x = 0$ (top) and $q_y = 0$ (bottom).
The Pei-Huang \cite{Pei12} triangular lattice of air holes
$r_\mathrm{cyl} = 0.42a$, $\epsilon_{\mathrm{host}} = 12.25$, $a = 0.365 \lambda$.
The circular line indicates the boundary of the hole.}
\label{fig:abs-E-Bloch-mode-triang-lattice-a0365-qx0-qy0}
\end{figure}

For this structure, we have calculated the Bloch bands and modes, as well
as wave transmission and reflection. All of these simulations employed
high-order FLAME difference schemes
\cite{tsukerman_05_1,tsukerman_06_1,Tsukerman-book07,Tsukerman-PBG08}.
These schemes do not necessarily operate on Cartesian grids and, in particular,
have been adapted to rhombic ones for the calculation of Bloch modes.
Selected results follow.

The first-Brillouin-zone equal frequency contour for $a = 0.365 \lambda$ 
($\lambda$ being the vacuum wavelength), close to the $\Gamma$-point
$a \approx 0.368 \lambda$, is shown in \figref{fig:circular-isofreq-triang-lattice-a0365}. 
The contour is indeed seen to be almost circular.
%

Two modes with $q_x = 0$ and $q_y = 0$ are plotted in \figref{fig:abs-E-Bloch-mode-triang-lattice-a0365-qx0-qy0}.
Incidentally, in contrast with rectangular lattices, for hexahedral ones the 
Bloch mode with $q_y = 0$ is \textit{not} generally lattice-periodic in the 
$y$-direction. Indeed, consider a point $\bfr$ on the lower side of the
rhombic cell and the corresponding point $\bfr -\frac12 \bfa_1 + \bfa_2$ on the upper side. 
For a plane-wave component $(m_1, m_2)$ of a Bloch wave, the respective 
phase factor between the two points is equal to unity only for even values
of $m_1$:
$$
  \exp(-i \, \frac{m_1}{2} \bfb_1 \cdot\bfa_1) \,
  \exp(i m_2 \bfb_2 \cdot \bfa_2)  
$$
$$
  =\,  \exp(-\pi i m_1) \, \exp(2\pi i m_2) ~=~ (-1)^{m_1}
$$
where the Kronecker-delta property of the lattice vectors was taken into account.

However, a similar calculation shows that for $q_x = 0$ lattice periodicity
in the $x$ direction does hold. Indeed, in that case the phase factor is
$$
  \exp(i m_1 \bfb_1 \cdot \bfa_1)
  \exp(i m_2 \bfb_2 \cdot \bfa_1)
$$
$$
    =\, \exp(i 2\pi m_1) \exp(i 2\pi m_2) \,=\, 1
$$

Numerical results in the following subsection illustrate that the
uncertainty principle is valid even for this highly isotropic case.

\begin{figure}
\centering
\includegraphics[width=0.99\linewidth]{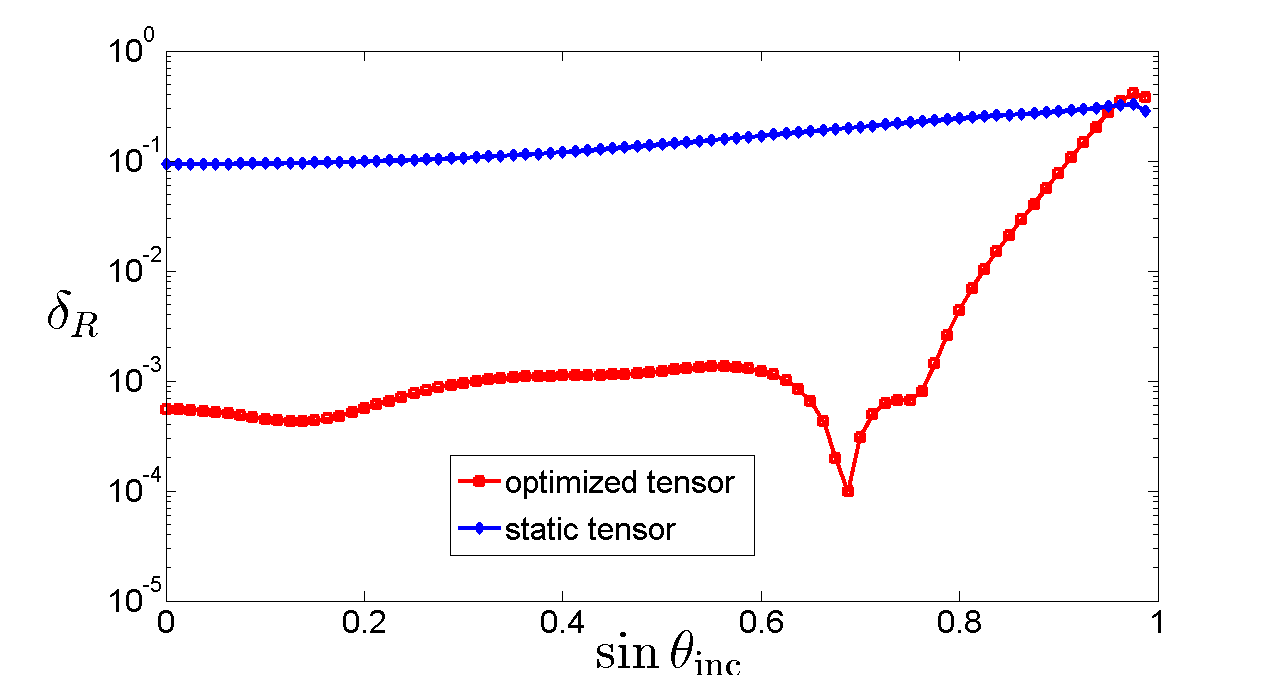}
\vskip 0.2in
\includegraphics[width=0.99\linewidth]{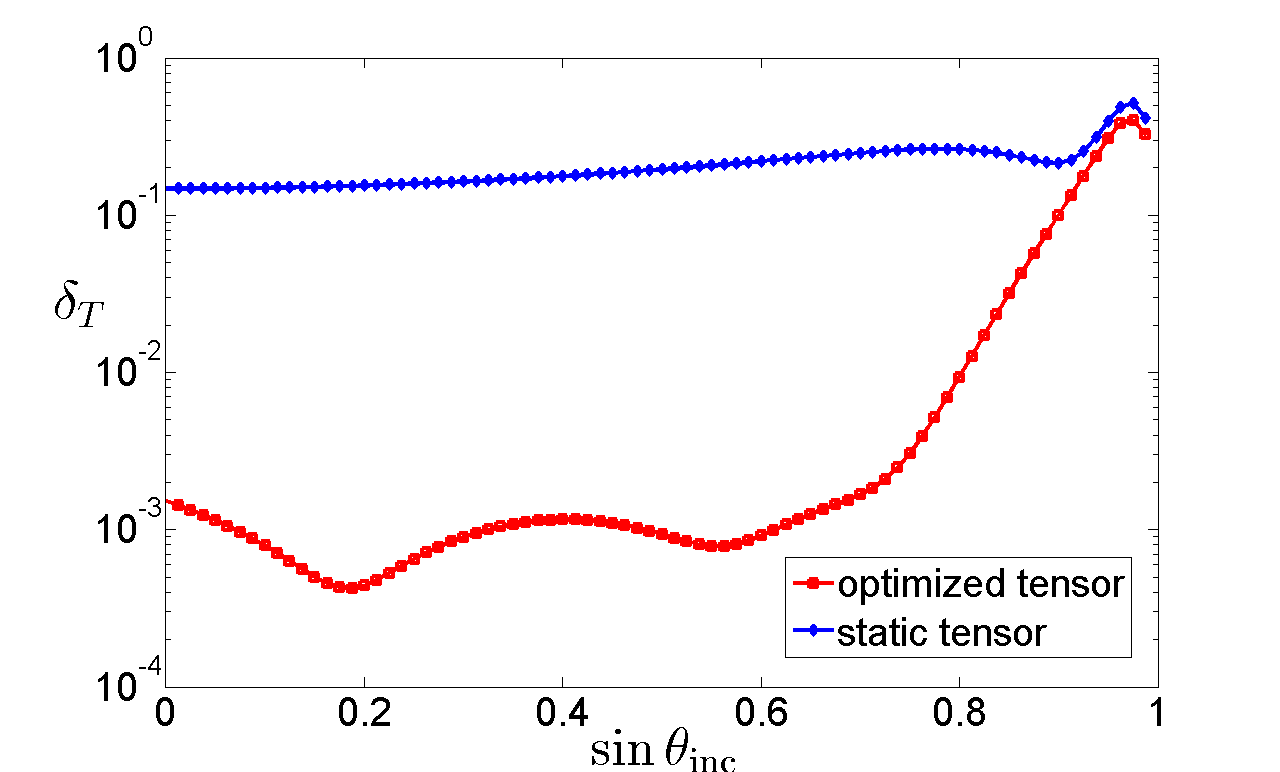}
\caption{Absolute errors in $R$ (top) and $T$ (bottom) as a function of the
sine of the angle of incidence. The errors are defined as
$\delta_R(\thinc) = |R_{\mathrm{opt}}(\thinc)  - R_{\mathrm{FD}}(\thinc)|$, 
$\delta_T(\thinc) = |T_{\mathrm{opt}}(\thinc) - T_{\mathrm{FD}}(\thinc)|$,
where `opt' refers to the optimized effective tensor and `FD' --
to accurate finite difference (FLAME) simulations. Tensor optimization was performed
within the range $[0, \pi/4]$ for the angle of incidence. $a / \lambda = 0.1$.}
\label{fig:error-RT-vs-theta-hex-lattice-a01-maxangle-pi4}
\end{figure}

\begin{figure}
\centering
\includegraphics[width=0.99\linewidth]{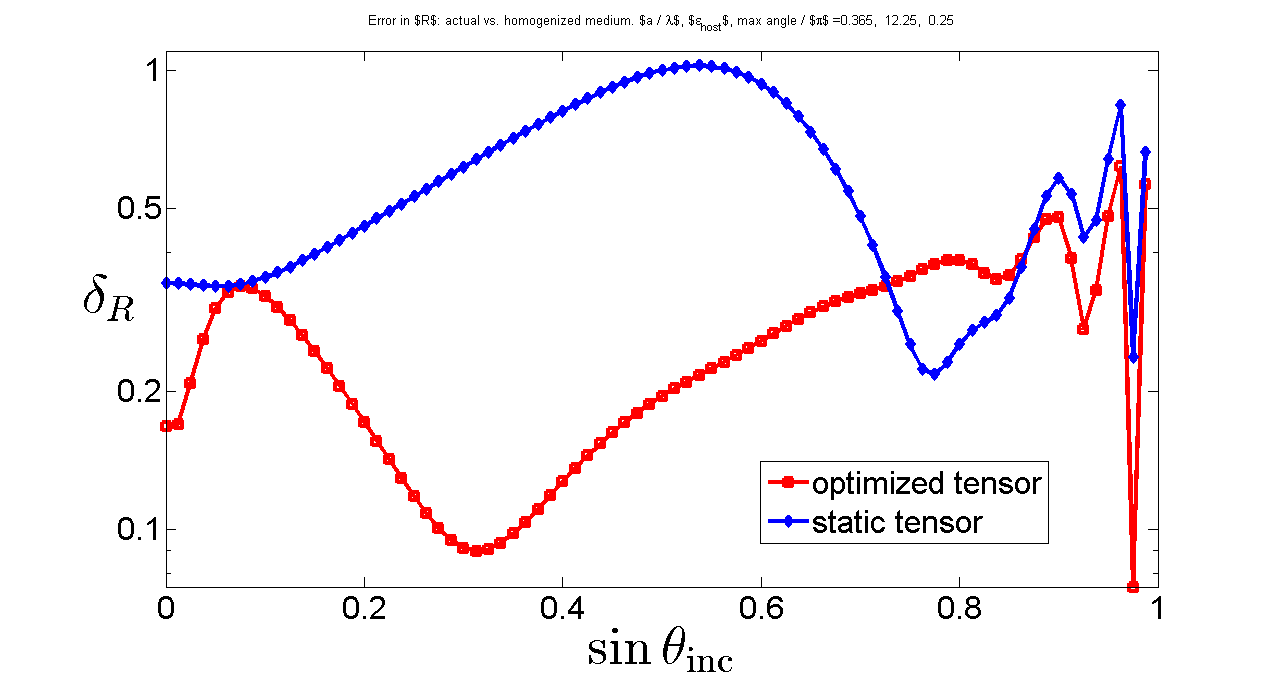}
\vskip 0.2in
\includegraphics[width=0.99\linewidth]{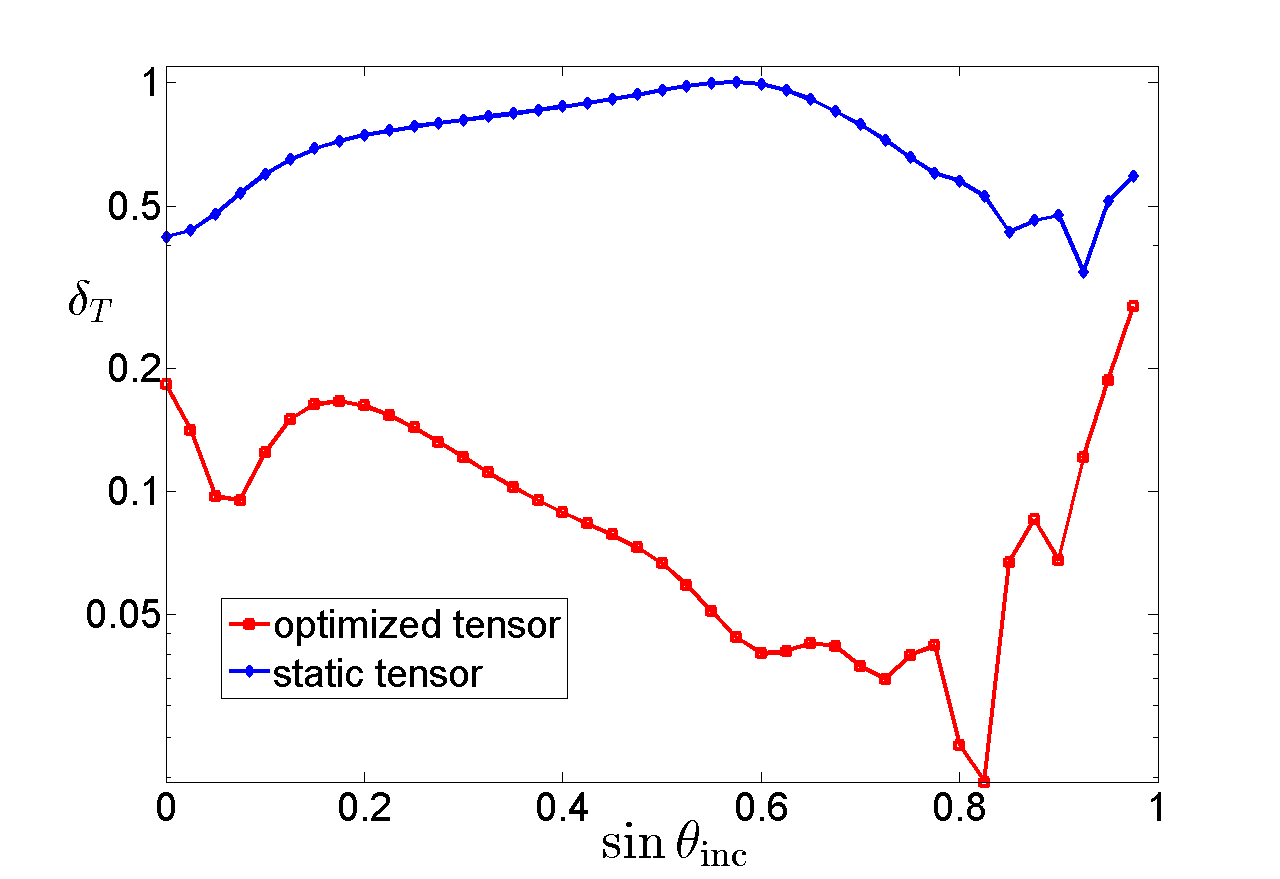}
\caption{Same as \figref{fig:error-RT-vs-theta-hex-lattice-a01-maxangle-pi4}
but for $a / \lambda = 0.365$. Note the logarithmic scale on the vertical axis.}
\label{fig:error-RT-vs-theta-hex-lattice-a0365-maxangle-pi4}
\end{figure}

%
\subsection{An Optimized Tensor}\label{sec:Optimal-tensor}
%
To verify the uncertainty principle numerically, we performed
``brute-force'' minimization of the TR-discrepancy with respect to 
a varying effective material tensor. The TR data from accurate FD simulations 
of wave propagation through a hexahedral-lattice slab was taken as a basis.
Parameters of the lattice are given in the previous subsection;
the angle of incidence varied from zero to an adjustable value $\theta_{\max}$. 
The number of layers in the slab was fixed at eight, which, 
according to extensive published evidence, should yield a very reasonable
representation of bulk behavior.
The Matlab$^\circledR \,$ optimization function {\tt fminsearch} 
was run repeatedly from different initial guesses for the tensor.
({\tt fminsearch} employs the Nelder-Mead simplex 
search method that does not use numerical or analytic derivatives.)
Included in our optimization routine was a simulated annealing
procedure \cite{Kirkpatrick83,Otten89} which allows the numerical solution to
escape from a local minimum, in the search for a global one.
Admittedly, in complex nonlinear optimization it can almost never be claimed
(with the notable exception of convex problems) that a global minimum
has been found. Still, in our case minimization was surprisingly
robust and converged to the same final result regardless
of the initial guess. 

This tensor optimization can be viewed as a generalization of
the traditional S-parameter retrieval, except that illumination is
not limited to normal incidence and the material tensor is not limited to
a diagonal one. Optimization was performed under constraints
proved formally in Appendix B: namely, the magnetoelectric coupling entries
of the tensor are purely imaginary, while all remaining ones are real.

For reference, we also consider the static tensor corresponding to
the $a / \lambda \rightarrow 0$ limit. This tensor is diagonal,
its magnetic part being the identity tensor. For $s$ polarization,
the effective static-limit permittivity is just the volume average
of those of the host and inclusion.

Figs.~\ref{fig:error-RT-vs-theta-hex-lattice-a01-maxangle-pi4} and
\ref{fig:error-RT-vs-theta-hex-lattice-a0365-maxangle-pi4} contrast the accuracy
of homogenization for the cases of a long wavelength ($a / \lambda = 0.1$)
and a short one ($a / \lambda = 0.365$), in correlation
with the corresponding magnetic effects. As already noted,
$a / \lambda = 0.365$ is close to the second $\Gamma$-point.

For the long wavelength (\figref{fig:error-RT-vs-theta-hex-lattice-a01-maxangle-pi4}),
one observes that the optimized tensor yields a good ``engineering level'' of accuracy:
the TR errors are below 0.01 in a broad range of (albeit not all) angles of illumination.
Even the static tensor in that case is borderline acceptable. This does not violate the uncertainty principle
for TR, as magnetic effects in the long-wavelength case are weak: $\mu \approx 1.04$
for the optimized tensor and $\mu = 1$ for the static one.

The situation is completely different for the short wavelength near the second
$\Gamma$-point (\figref{fig:error-RT-vs-theta-hex-lattice-a0365-maxangle-pi4}).
Not surprisingly, the static tensor in this regime is not applicable at all.
But even the optimized tensor does not work: the errors are too large,
except for an accidental narrow range of the angles of incidence.
(Clearly, for any given specific angle, TR can be represented perfectly
just by parameter fitting.)

\section{Conclusion}\label{sec:Conclusion}

The paper demonstrates that a nontrivial effective permeability tensor of periodic structures 
composed of intrinsically nonmagnetic constituents has limitations and is subject
to (at least) two ``uncertainty  principles''. First, the stronger the magnetic response 
(as measured by the deviation of the optimal effective permeability tensor from identity),
the less accurate (``certain'') predictions of the effective medium theory. 
Second, also in the case of a strong magnetic response, homogenization cannot 
simultaneously and accurately reproduce both TR \textit{and} power relations in the periodic structure.
In practice, there is still room for engineering design, but the trade-offs
between magnetic response and the accuracy of homogenization must be noted.

These conclusions follow from the analysis of coarse-level fields that must
satisfy the dispersion relation and boundary conditions accurately, while simultaneously
approximating the far field reflected and transmitted by a metamaterial sample.
All of this implies that not only the dispersion relation but also
surface impedance have to be illumination-independent. These prerequisites cannot
unfortunately be simultaneously satisfied if the desired magnetic response is strong.
As a supporting example, we considered a hexahedral lattice of cylindrical air holes
in a dielectric host, as investigated previously in Ref.~\onlinecite{Pei12}.
Even in this highly isotropic case, seemingly conducive to homogenization,
the uncertainty principles remain valid.

\section*{Acknowledgment}
The work of Vadim A. Markel has been carried 
out thanks to the support of the A*MIDEX project 
(No. ANR-11-IDEX-0001-02) funded by the ''Investissements d'Avenir'' 
French Government program, managed by the French National Research Agency (ANR).
The research of both authors was also supported by the US National Science Foundation
under Grant DMS-1216970. 

\section*{Appendix A: Dependence of Bloch Waves on Direction of Propagation}

As no analytical expressions
are available for Bloch waves, let us consider an approximation and examine its
implications for the magnetic parameter $\zeta$ in \eqnref{eqn:zeta-tau-via-e-zeromean}.
Since \eqnref{eqn:zeta-tau-via-e-zeromean} is valid for the case of a four-fold
symmetry, we continue to operate under that assumption. To simplify analytical manipulations,
we assume, in addition, that the metamaterial consists of $C_4$-symmetric
but otherwise arbitrarily shaped particles embedded in a homogeneous host; 
then the medium next to the lattice cell boundary is homogeneous, which is
quite typical.
Finally, we shall continue to concentrate on the 2D case, $s$-mode; it will be
clear from the analysis below that this assumption is not critical, but it does
simplify the mathematics greatly.

Under the above assumptions, the $e$-field at the cell boundary
can be expanded into cylindrical harmonics:
\begin{equation}\label{eqn:cyl-harmonic-expansion-inc-scattered}
   e(\bfr) \,=\, \sum_{m=-\infty}^{\infty} \left[ c_m J_m(kr) + s_m h_m(kr) \right]
   \, \exp(\iu m \phi)
\end{equation}
where $J_m$ and $h_m$ are the Bessel function and the Hankel function of the first kind,
respectively; $c_m$ and $s_m$ are (yet undetermined) coefficients, and $k$ is
the wavenumber corresponding to the host material around the cell boundary.
The (infinite) coefficient vectors $\underline{c} = (\ldots, c_{-1}, c_0, c_1, \ldots)$
and $\underline{s} = (\ldots, s_{-1}, s_0, s_1, \ldots)$ are linearly related:
\begin{equation}\label{eqn:s-vs-c-via-T}
   \underline{s} = T \underline{c}
\end{equation}
Eq. \eqnref{eqn:s-vs-c-via-T} may serve as a definition of the scattering matrix
$T$ which depends on the particle in the cell and fully characterizes its
electromagnetic response.

As an approximation, let us retain the terms in \eqnref{eqn:cyl-harmonic-expansion-inc-scattered}
up to quadrupole ($|m| \leq 2$); the coefficient vectors then reduce to length five,
and $T$ is $5 \times 5$.

Since the Bloch wave is defined up to an arbitrary scaling factor, we need
four conditions to fix the coefficient vectors $\underline{c}$ and $\underline{s}$.
The simplest way to impose such conditions is by collocation at the four edge
midpoints of the boundary.

More precisely, let the square lattice cell $a \times a$ be centered at the origin;
let midpoints 1 and 2 correspond to the bottom and top edges, respectively:
$\bfr_1 = (0, -a/2)$, $\bfr_2 = (0, a/2)$. We require that the fields at these midpoints be
related by the Bloch condition
\begin{equation}\label{eqn:e2-vs-e1-via-Bloch}
   e(\bfr_2) - \lambda_n e(\bfr_1) = 0; ~~~ \partial_n e(\bfr_2) - \lambda_n \partial_n e(\bfr_1) = 0
\end{equation}
where
$
   \lambda_n \equiv \exp(\iu aq \cos \theta_B)
$.
In a completely similar manner, for the midpoints on the ``left''
and ``right'' edges, $\bfr_3 = ( -a/2, 0)$, $\bfr_4 = (a/2, 0)$, the Bloch condition is
\begin{equation}\label{eqn:e4-vs-e3-via-Bloch}
   e(\bfr_4) - \lambda_{\tau} e(\bfr_3) = 0; 
   ~~~ \partial_{\tau} e(\bfr_4) - \lambda_{\tau} \partial_{\tau} e(\bfr_3) = 0
\end{equation}
where
$
   \lambda_{\tau} \equiv \exp(\iu aq \sin \theta_B)
$.
It is straightforward to write the midpoint collocation condition in
matrix-vector form
\begin{equation}\label{eqn:Bloch-collocation-condition-1}
  \left( \mathcal{J}_{\alpha+1} - \lambda_n \mathcal{J}_{\alpha} \right) \underline{c} 
  \,+\, \left( \mathcal{I}_{\alpha+1} - \lambda_n \mathcal{I}_{\alpha} \right) \underline{s}  = 0,
\end{equation}
\begin{equation}\label{eqn:Bloch-collocation-condition-2}
  \left( \mathcal{J}_{\alpha+1}^{\partial} - \lambda_n \mathcal{J}_{\alpha}^{\partial} \right) \underline{c} 
  \,+\, \left( \mathcal{I}_{\alpha+1}^{\partial} - \lambda_n \mathcal{I}_{\alpha}^{\partial} \right) \underline{s}  = 0,
\end{equation}
$\alpha = 1,3$.

%
All $\mathcal{J}$s and $\mathcal{I}$s above are row vectors of length five.
Vector $\mathcal{J}_l$ contains the values of the Bessel functions at the
collocation point $l$ -- that is, the values $J_m(\bfr_l)$, $m = 0, \pm 1, \pm2$.
Likewise, vector $\mathcal{I}_l$ contains the values of the respective Hankel functions
$h_m(\bfr_l)$. Vectors labeled with superscript $\partial$ are analogous
but contain the respective partial derivatives of the Bessel/Hankel functions:
$\partial_n$ for collocation points 1 and 2, and $\partial_{\tau}$ for points 3 and 4.

Recalling now that $\underline{s} = T \underline{c}$ \eqnref{eqn:s-vs-c-via-T}
and merging the four conditions above into a single matrix, we have
\begin{equation}\label{eqn:c-eq-Null-A}
   \underline{c} ~=~ \mathrm{Null} \, A
\end{equation}
where
\begin{equation}\label{eqn:Bloch-collocation-conditions-merged}
   A ~=~
   \begin{pmatrix}
   \mathcal{J}_{2} + T \mathcal{I}_{2} - \lambda_n (\mathcal{J}_{1} + T \mathcal{I}_{1}) \\
   \mathcal{J}_{2}^{\partial} + T \mathcal{I}_{2}^{\partial} 
               - \lambda_n (\mathcal{J}_{1}^{\partial} + T \mathcal{I}_{1}^{\partial}) \\
   \mathcal{J}_{4} + T \mathcal{I}_{4} - \lambda_{\tau} (\mathcal{J}_{3} + T \mathcal{I}_{3}) \\
   \mathcal{J}_{4}^{\partial} + T \mathcal{I}_{4}^{\partial} 
               - \lambda_{\tau} (\mathcal{J}_{3}^{\partial} + T \mathcal{I}_{3}^{\partial}) \\
   \end{pmatrix}
\end{equation}
Now that the Bloch wave expansion into cylindrical harmonics has been
evaluated, we can substitute it into expression \eqnref{eqn:zeta-tau-via-etilde-dn}
for $\zeta_{\tau \tau}$:
\begin{equation}\label{eqn:zeta-tau-matrix-form}
   \zeta_{\tau \tau} \,\approx\, 
    \frac{\iu}{q \cos \theta_B} \,
   \frac{(\mathcal{J}_{1}^{\partial} + T \mathcal{I}_{1}^{\partial}) \underline{c} } 
   {(\mathcal{J}_{1} + T \mathcal{I}_{1}) \underline{c}}
\end{equation}
The key point here is that coefficients $\underline{c}$ depend in quite a convoluted way
on the angle. Indeed, vector $\underline{c}$ is the null space of matrix $A$
which contains $\lambda_{\tau}$ and $\lambda_{\tau}$, which in turn are 
complex exponentials of $\cos \theta_B$ and $\sin \theta_B$.
This convoluted angular dependence of $\underline{c}$ translates, via 
\eqnref{eqn:zeta-tau-matrix-form}, into an even more complex angular dependence
of $\zeta_{\tau \tau}$.

\section*{Appendix B: Properties of the Tensor, s-mode}

This section includes a formal proof of some properties of
the optimized material tensor, under natural assumptions about
this optimization. The general plan of analysis is as follows:
\begin{enumerate}
\item Assume some valid fields $\bfe, \bfh, \bfd, \bfb$ in and around a metamaterial slab.
\item  Apply a transformation (``symmetry'') $\calS$ 
with regard to which Maxwell's equations are invariant: $e' = \calS e$, etc.
\item Find coarse-level $\bfE, \bfH, \bfD, \bfB$:
       $\bfE = f_1(\bfe)$,  $\bfH = f_2(\bfh)$, $\bfD = g_1(\bfH)$, $\bfB = g_2(\bfE)$;
       $\bfE' = f_1(\bfe')$, etc.,
    where functions $f_{1,2}$ are boundary averages\cite{Tsukerman-Markel14} 
    and functions  $g_{1,2}$ come from Maxwell's equations.
\item Given
        $\{\bfD, \bfB\} = \calM \{\bfE, \bfH\}$,
        $\{\bfD', \bfB'\} = \calM \{\bfE', \bfH'\}$
     determine the implications for $\calM$.
\end{enumerate}     

Let us implement this plan if $\calS$ is complex conjugation.
The governing equation for the $s$-mode is
\begin{equation}\label{eqn:del2-e-k2-eps-e}
  \nabla^2 e(\bfr) + \K^2 \epsilon(\bfr) e(\bfr) ~=~ 0
\end{equation}
This equation is indeed invariant with respect to complex conjugation
$\calS$ if $\epsilon$ is real.
The original and transformed Bloch waves are, for a given $\bfq$,
\begin{equation}\label{eqn:e-Bloch-wave-copy}
  e_B(\bfr) ~=~ \tilde{e}_B(\bfr) \exp(\iu \bfq \cdot \bfr) 
\end{equation}
\begin{equation}\label{eqn:h-Bloch-wave-copy}
  h_B(\bfr) \,=\, \frac{1}{\iu \K} \frac{\partial e_B}{\partial n}
  \,=\, 
  \tilde{h}_{B}(\bfr) \exp(\iu \bfq \cdot \bfr),
\end{equation}
(tangential component). Here
\begin{equation}\label{eqn:tilde-h-via-e-copy}
  \tilde{h}_{B}(\bfr) \,=\, \frac{q_n}{\K} \, \, \tilde{e}_B(\bfr) + 
  \frac{1}{\iu \K} \frac{\partial \tilde{e}_B(\bfr)}{\partial n}
\end{equation}
For a real $\bfq$,
\begin{equation}\label{eqn:e-conj-Bloch-wave}
  e'_B(\bfr) ~=~ \tilde{e}_B^*(\bfr) \exp(-\iu \bfq \cdot \bfr) 
\end{equation}
with
\begin{equation}\label{eqn:h-prime-Bloch-wave-copy}
  h'_B(\bfr) \,=\, \frac{1}{\iu \K} \frac{\partial e'_B}{\partial n}
  \,=\, 
  \tilde{h}'_{B}(\bfr) \exp(-\iu \bfq \cdot \bfr),
\end{equation}
\begin{equation}\label{eqn:tilde-h-via-e-prime}
  \tilde{h}'_{B}(\bfr) \,=\, -\frac{q_n}{\K} \, \, \tilde{e}^*_B(\bfr) + 
  \frac{1}{\iu \K} \frac{\partial \tilde{e}^*_B(\bfr)}{\partial n}
  ~=~ -\tilde{h}_{B}^*(\bfr)
\end{equation}
The above derivation formally shows that, as could be expected, 
if $\{\bfe(\bfr, \bfq), \bfh(\bfr, \bfq)\}$ is a valid Bloch wave in a lossless structure,
then $\{\bfe^*(\bfr, -\bfq), -\bfh^*(\bfr, -\bfq)\}$ is also a valid Bloch wave.
(Notably, the sign of the magnetic field is reversed as the direction
of the wave is reversed.)

The amplitudes of the two respective plane waves in the homogenized medium
are therefore related as follows:
\begin{equation}\label{eqn:E0-prime-eq-E0-conj}
   E'_0 \,=\, \langle \tilde{e}^*_B \rangle_S \,=\, E_0^*
\end{equation}
\begin{equation}\label{eqn:H0-prime-eq-minus-H0-conj}
   H'_0 \,=\, \langle \tilde{h}'_B \rangle_S \,=\, -\langle \tilde{h}^*_B \rangle_S \,=\, -H_0^*
\end{equation}
\begin{equation}\label{eqn:B0-prime-eq-minus-B0-conj}
   B'_{0} \,=\, -\frac{q_{n}}{\K} \, \langle \tilde{e'}_B \rangle_S 
   \,=\, -\frac{q_{n}}{\K} \, E_0^* \,=\, -B_0^*
\end{equation}
Let there be a material tensor $\calM$ that relates the fields as follows:
\begin{equation}\label{eqn:calM-EH-eq-DB}
   \begin{pmatrix}
   \calM_{11} & \calM_{12}  & \calM_{13}\\
   \calM_{21} & \calM_{22}  & \calM_{23}\\
   \calM_{31} & \calM_{32}  & \calM_{33}
   \end{pmatrix} \,
     \begin{pmatrix}
     E_0 \\ H_{0x} \\ H_{0y}
     \end{pmatrix}
     ~=~
     \begin{pmatrix}
     D_0 \\ B_{0x} \\ B_{0y}
     \end{pmatrix}
\end{equation}
Then from \eqnref{eqn:E0-prime-eq-E0-conj} -- \eqnref{eqn:B0-prime-eq-minus-B0-conj}
it follows that entries $\calM_{12}$, $\calM_{13}$, $\calM_{21}$, $\calM_{31}$  are purely imaginary,
while all others are real. Indeed, substituting into \eqnref{eqn:calM-EH-eq-DB} 
a valid wave with amplitudes $(E_0, H_{0x}, H_{0y})$,
we have, say, for the first equation in the system
$$
 \calM_{11} E_0 \,+\, \calM_{12} H_{0x} + \calM_{13} H_{0y} ~=~ D_0
$$
and for the corresponding conjugate wave with $(E_0^*, -H_{0x}^*, -H_{0y}^*)$,
$$
 \calM_{11} E_0^* \,-\, \calM_{12} H_{0x}^* - \calM_{13} H_{0y}^* ~=~ D_0^*
$$
From the above equations for the two waves, it immediately follows  that
$$
 \left[ (\calM_{11} - \calM_{11}^*)  \,+\, (\calM_{12} + \calM_{12}^*) \, \eta_x(\theta_B)
   \right.
$$
\begin{equation}\label{eqn:calM-calM-star-eq-0}  
   \left. 
   + \, (\calM_{13} + \calM_{13}^*) \, \eta_y(\theta_B) \right] ~=~ 0
\end{equation}
where we introduced the notation $\eta_{x,y} \equiv E_0 / H_{0x,0y}$
and noted that the $\eta$s depend on the direction of propagation
of the Bloch wave. Eq. \eqref{eqn:calM-calM-star-eq-0} can hold
for all directions of propagation only if
$$
  \calM_{11} = \calM_{11}^*, ~~ \calM_{12} =-\calM_{12}^*,
  ~~   \calM_{13} = -\calM_{13}^*,
$$
i.e. if $\calM_{11}$ is real and $\calM_{12}$, $\calM_{13}$ are purely imaginary.
Similarly, $\calM_{21}$, $\calM_{31}$ must also be purely imaginary,
while $\calM_{22}$, $\calM_{23}$ $\calM_{32}$, and $\calM_{33}$ must be real.

\bibliographystyle{plain}
\bibliography{abbrev,book,master,Igor_reference_dbase,local}

\end{document}